Periodicity-dependence of the ferroelectric properties in $BiFeO_3$/$SrTiO_3$

multiferroic superlattices

R. Ranjith, B. Kundys and W. Prellier[1]

*Laboratoire CRISMAT, CNRS UMR 6508, ENSICAEN*

*6 Bd Maréchal Juin, 14050 Caen Cedex, France.*

Artificial superlattices of $(BiFeO_3)_m(SrTiO_3)_m$ (m= 1 to 10 unit cells) consisting of multiferroic $BiFeO_3$ and insulating $SrTiO_3$ layers were fabricated on (100)-oriented $SrTiO_3$ substrates by pulsed laser ablation. The remnant polarization and leakage current behavior were studied varying the periodicity (8-80Å) of the superlattice. The leakage current was reduced by few orders of magnitude on increase of periodicity compared to single layer $BiFeO_3$ thin films. Reduced leakage and intrinsic polarization hysteresis was observed and was confirmed by PUND analysis for periodicities in the range ~ 20-60 Å. The leakage current was observed to be dominated by space charge limited conduction.

---

[1] prellier@ensicaen.fr

Multiferroic - a material that posses both ferroelectricity and (ferro) magnetism, has been a potential candidate for fundamental studies and technological applications like striction based devices and memory applications.[1,2] $BiFeO_3$ has been one among the few naturally existing multiferroic materials that has been studied extensively in the past few decades, both in the form of bulk and thin films.[3] $BiFeO_3$ possess a magnetic Neel temperature of ~673K and a ferroelectric Curie temperature of ~1093K in the bulk form.[3] However, there are few practical hurdles in utilizing the material for device applications despite its interesting properties. In particular, the large ferroelectric coercive field and high leakage behavior are considered as the major draw backs for $BiFeO_3$ thin films.[4,5] Due to high sensitivity of the $BiFeO_3$ thin films on the process parameters, the leakage behavior of $BiFeO_3$ has been attributed to the presence of parasitic phases like bismuth oxide and iron oxide.[6] The presence of oxygen vacancies and the transformation of a fraction of $Fe^{3+}$ cation to $Fe^{2+}$ cation are attributed to the leakage behavior of the $BiFeO_3$ thin films fabricated without any parasitic phases.[4,5] As a consequence, many studies are devoted to the reduction of this leakage and understanding the leakage mechanisms in $BiFeO_3$ thin films.[7,8,9,10,11,12] In epitaxially grown $BiFeO_3$ thin films, the leakage mechanism is dominated by Poole – Frenkel (P-F) emission with symmetric electrode configuration, while a mechanism close to (P-F) emission in an asymmetric electrode configuration[4]. Various efforts like i) substitution of heterovalent and homovalent substitutions on both A and B site cations which effectively controls the anion vacancies,[5,7-11]ii) modification of the electrode material interface[12] and iii) recently a superlattice structure of a single periodicity of $BiFeO_3$ with $SrTiO_3$[13] were thusly employed to reduce the leakage behavior of $BiFeO_3$ thin films. Doping cations like, La,

Mn, Ni, Ti, Nb and Sc in both the $Bi^{3+}$ and $Fe^{3+}$ sites has been studied extensively.[5,7-11] Among the $A$ site substitution, La doping was found to enhance the polarization and reduce the leakage behavior of $BiFeO_3$ thin films.[7] Substitution of cations like $Sc^{3+}$, $Nb^{5+}$, $Cr^{3+}$, $Mn^{3+}$ in the $B$ site, $Sc^{3+}$ and $Nb^{5+}$ were also found to improve the electrical properties.[8-11] The leakage current increases with the $Ni^{3+}$-doping and decreases with the $Ti^{4+}$-doping.[5] The creation and annihilation of oxygen defects and the alteration of anion vacancies due to chemical heterogeneity was attributed to the leakage current in $BiFeO_3$ thin films.[5] Artificial superlattices provide excellent opportunities to manipulate the strain and the chemical heterogeneity in order to exhibit completely new or enhance the properties, which were absent in the parent compounds.[14] Our motivations are as follows: $SrTiO_3$, a room temperature paraelectric insulator, with a close lattice match with $BiFeO_3$ could be a useful candidate to introduce strain and chemical heterogeneity. The periodicity of a given superlattice structure is crucial on tailoring the collective physical properties of the system. In addition, there have been no extensive studies reported on the periodicity dependence of the electrical behavior of $(BiFeO_3)/(SrTiO_3)$-based superlattices.

In this letter, a series of artificial superlattices structures made of $BiFeO_3$ and $SrTiO_3$ layers was fabricated using the pulsed laser deposition technique. The room temperature remnant polarization ($P_r$), leakage current density (J) of the $(BiFeO_3)_m(SrTiO_3)_m$ (m=1 to 10 unit cells) superlattices and their periodicity dependence are studied. Remnant polarization and leakage current density displays an optimum for a periodicity range of ($\sim$ 20-60 Å) and a possible mechanism of the leakage current in these superlattices structures is proposed.

Thin films of $BiFeO_3$ and $SrTiO_3$ and their superlattices were grown on (001) oriented $SrTiO_3$ substrates(CrysTec, Germany), at $700^oC$ in oxygen ambient of 20 mTorr using a multitarget pulsed laser ablation and deposition technique. A pulsed KrF excimer laser ($\lambda$=248 nm) was fired onto the ceramic targets at a repetition rate of 3 Hz. The deposition rates (typically ~ 0.1 Å/pulse) of $BiFeO_3$ and $SrTiO_3$ were calibrated individually for each laser pulse of energy density ~ 1.5 $J/cm^2$. On completion of deposition the chamber was filled with oxygen ambient of 300 mTorr on cooling upto $500^oC$ and flushed with oxygen pressure of 300 Torr and cooled down further to room temperature at the rate of 5°C/min. The superlattice structures were synthesized by repeating several times the bilayer consisting of m unit cells thick $BiFeO_3$ layer and m unit cells thick $SrTiO_3$ layer, with m taking integer values from 1 to 10, keeping a constant total thickness of the superlattice equal to 2400Å. A series of superlattices with periodicity ranging from ~8-80Å were thusly prepared. Prior to the growth of superlattice, a bottom electrode of $LaNiO_3$ (800Å) was deposited at $700^oC$ in oxygen ambient of 100 mTorr. Gold pads of 400x400$\mu m^2$ sized (physical mask) were sputtered on top of the superlattice structures and on top of $LaNiO_3$ regions unexposed to the superlattice deposition. The fabricated heterostructures were characterized in the metal-insulator-metal (MIM) configuration to study their electrical properties. Ferroelectric polarization hysteresis (P-E) and PUND measurements were carried out on a Radiant Technology high precision loop tester and the dc leakage characteristics were studied using a Keithley electrometer (6517A). The experimental details of the PUND analysis could be found elsewhere.[4]

In the conventional $\Theta-2\Theta$ scans of the superlattices, no peaks were observed other than the (00l) Bragg reflections of the electrode, the constituents, the substrate and the satellites due to chemical modulation present in the multilayer. A typical example is given as inset of Figure 1 for a superlattice with a periodicity of ~40Å (i.e. made from 5 u.c. of $BiFeO_3$ and 5 u.c. of $SrTiO_3$). The denoted number i indicates the $i^{th}$ satellite peak. As the periodicity varies, the fundamental 001 diffraction peak of the constituents shifted. The presence of second order strong satellite peaks on either side of the fundamental (001) diffraction clearly show the formation of a new structure having a periodic chemical modulation of the constituents. The periodic chemical modulation ($\Lambda$) for the superlattice, as extracted from $\Theta-2\Theta$ XRD, is in close agreement with theoretical values (~ 40 Å) based on the lattice parameters of each constituent (5 x 4.010 Å for $BiFeO_3$ and 5 x 3.902 Å for $SrTiO_3$). Figure 1 shows the variation of the average lattice spacing ($2xd_{002}=\Lambda/2m$) for a series of superlattices having various periodicity. The epitaxially grown phase pure $BiFeO_3$ thin films on $LaNiO_3$ deposited $SrTiO_3$ substrates exhibited an out-of-plane lattice parameter of ~ 4.010Å in close agreement with previous reports.[13] The periodicity-dependence of the average lattice parameter shown in Figure 1, reveals that the $BiFeO_3$ lattice is subjected to a compressive strain while the counterpart $SrTiO_3$ lattice is under a tensile strain. At low periodicities, the lattice could accommodate the strain and hence, the average lattice spacing ($\Lambda/2m = 2$ x $d_{002}$) is close to the single layer $BiFeO_3$ thin film on $LaNiO_3$. On increasing the periodicity the strain relaxes and a shift in the average lattice spacing towards the $SrTiO_3$ substrate is observed.

Figure 2 displays the variation of the room temperature remnant polarization ($P_r$) as a function of the periodicity of the superlattices, and the inset shows the ferroelectric polarization hysteresis (P-E loop) of a $(BiFeO_3)_5(SrTiO_3)_5$ superlattice having a periodicity of $\sim 40$ Å measured at different frequencies. The coercive field ($E_c$) of the superlattice structure is $\sim 78$kV/cm and is independent of measured frequency range of 1-10 kHz (inset of Figure 2). Figure 3 shows the $P_r$ observed from the PUND analysis carried out on the same structure. The frequency-independent behavior of the $P_r$ shows that the polarization observed is intrinsic to the superlattice structure and does not arise from the mobile charges or other extrinsic effects due to leakage current. The $P_r$ values observed from both the P-E and PUND measurements were consistent and hence confirm intrinsic FE behavior of the superlattice structures. The FE domains and the piezoelectric coefficients and their periodicity dependence on BFO-STO superlattices were observed through piezo force microscopy (PFM) and will be reported elsewhere. In $BiFeO_3$, ferroelectricity results from the local distortion introduced by the stereo-chemically active $6s^2$ lone pair electrons of the $Bi^{2+}$ cation.[15] Remnant polarization ($P_r$) values in the range of $2.5 - 5$ $\mu C/cm^2$ is observed within the periodicity range of $\sim 20$-60 Å, which is almost one order of magnitude less than reported values of $BiFeO_3$ films grown on $SrTiO_3$ substrate using a $SrRuO_3$ bottom electrode.[3] This low polarization could be attributed to different strain conditions experienced by the $BiFeO_3$ lattice due to the bottom electrode ($LaNiO_3$), the presence of the paraelectric $SrTiO_3$ layer and the possibility of charged interface between the $BiFeO_3$ and $SrTiO_3$ layers. In the case of $(BiFeO_3)_2(SrTiO_3)_2$, the leakage current was found dominating over the (P-E) measurements. On the other hand, the $(BiFeO_3)_{10}(SrTiO_3)_{10}$ superlattice with a periodicity of 80Å exhibited a higher $P_r$ value of $\sim 10$ $\mu C/cm^2$. The observed $E_c$ and $P_r$ values were dependent on frequency and applied field in the case of $(BiFeO_3)_{10}(SrTiO_3)_{10}$. Inspite of the low dc leakage, the (P-E) loop was dominated by the other extrinsic space charge effects, which could be due to the presence of charged interface, defects and misfit dislocations formed due to strain relaxation at the interface.[16] In addition to the polarization the leakage current densities of the superlattice structures were studied.

Figure 4 shows the room temperature leakage current density (J) as function of the superlattice periodicity. The current density exhibited a decreasing trend on increasing periodicity. The high symmetric nature of leakage current observed on reversal of applied field confirmed that the leakage current was not limited by the electrode material interface. The leakage current in ferroelectrics is known to be dominated by an electrode limited Schottky emission, or bulk limited Poole-Frenkel emission or space charge limited conduction (SCLC).[4,17] To further investigate the leakage mechanism the current-voltage (I-V) plots were fitted with different models. Inset of Figure 4 shows the (I-V) curve of a $(BiFeO_3)_{10}(SrTiO_3)_{10}$ ($\Lambda \sim 80$ Å). The observed I-V curve was analyzed with a Schottky plot, which is expected to show a linear behavior with the slope giving the high frequency dielectric constant which was not observed in the present case.[17] Consequently, the observed leakage behavior of the superlattice structures did not obey the Schottky type of conduction. The nature of the (I-V) curve shows that the leakage current could be due to the space charge limited conduction type of conduction. In the case of SCLC, a low field linear region extends to a certain cross over voltage ($V_{TFL}$) and beyond which, the current would rise suddenly and reach a saturation obeying a quadratic power law of I $\alpha$ $V^2$. The bulk generated charge carriers like oxygen vacancies could give rise to a linear (I-V) characteristic at low fields.[17,18] On exceeding $V_{TFL}$ the deep traps present in the system gets filled and beyond $V_{TFL}$ these charges gets detrapped and produces excess charges in the conduction band. Hence, before reaching the space charge limited current, a sudden jump is observed with a slope of 4.8 followed by the space charged limited region with a slope of 2.07. Three distinct regions are shown with a linear fitting along with their corresponding slopes in the inset of Figure 4. The formation of the Lampert triangle,[18] indicates that the leakage current of $(BiFeO_3)_m(SrTiO_3)_m$ superlattice is dominated by space charged limited conduction, whereas, a Poole-Frenkel emission type of conduction was observed in the single layer $BiFeO_3$.[4]

In summary, the leakage and the polarization behavior of a series of $(BiFeO_3)_m(SrTiO_3)_m$ superlattices deposited on (100)-oriented $SrTiO_3$ substrates were studied. The intrinsic remnant polarization of the superlattices was observed at the periodicities in the range of 20-60 Å. The observed polarization was consistent with the PUND analysis. Polarization measurements at lower (~10Å) and higher (~ 80 Å) periodicities were dominated by high dc leakage and extrinsic interface effects respectively. On increasing the $SrTiO_3$ layer thickness, the current density decreases and the leakage current is found to be space charge limited. Finally, the ferroelectric properties, namely leakage and intrinsic polarization, are optimum for a superlattice periodicities in the range of 20-60 Å (m=3-7) opening a route for improvement of multiferroic properties in devices using $BiFeO_3$ at room temperature.

This work was carried out in the frame of the NoE FAME (FP6-5001159-1), the STREP MaCoMuFi (NMP3-CT-2006-033221), and the STREP CoMePhS (NMP4-CT-2005-517039) supported by the European Community and by the CNRS, France. Partial support from the ANR (NT05-1-45177, NT05-3-41793) is also acknowledged. The authors would also like to acknowledge Dr. L. Mechin, Mr. C. Fur, Mr. J. Lecourt, Prof G. Poullain and Dr. Ch. Simon for their help in performing the experiments.

**Figure Captions**

Figure 1. (Color online) Average out-of-plane lattice parameter ($2 \times d_{002}$) as a function of the periodicity ($\Lambda$) for several superlattices. The line is a guide for the eyes. The inset shows a $\Theta-2\Theta$ pattern recorded around the 001 reflection of $SrTiO_3$ for a superlattice with a periodicity of ~40Å.

Figure 2. (Color online) Variation of remnant polarization ($P_r$) as a function of the periodicity ($\Lambda$) for several superlattices. The inset shows the room temperature (P-E) loop of a superlattice with $\Lambda \sim 40$Å measured at different frequencies. The line is drawn as a guide to the eyes.

Figure 3. Remnant polarization at different frequencies measured using PUND analysis for a superlattice with periodicity $\Lambda \sim 40$Å.

Figure 4. (Color online) Variation of leakage current density (J) as a function of the periodicity ($\Lambda$) for several superlattices. The inset shows room temperature (I-V) curve of superlattice with $\Lambda \sim 80$ Å. The slopes of the curve are indicated.

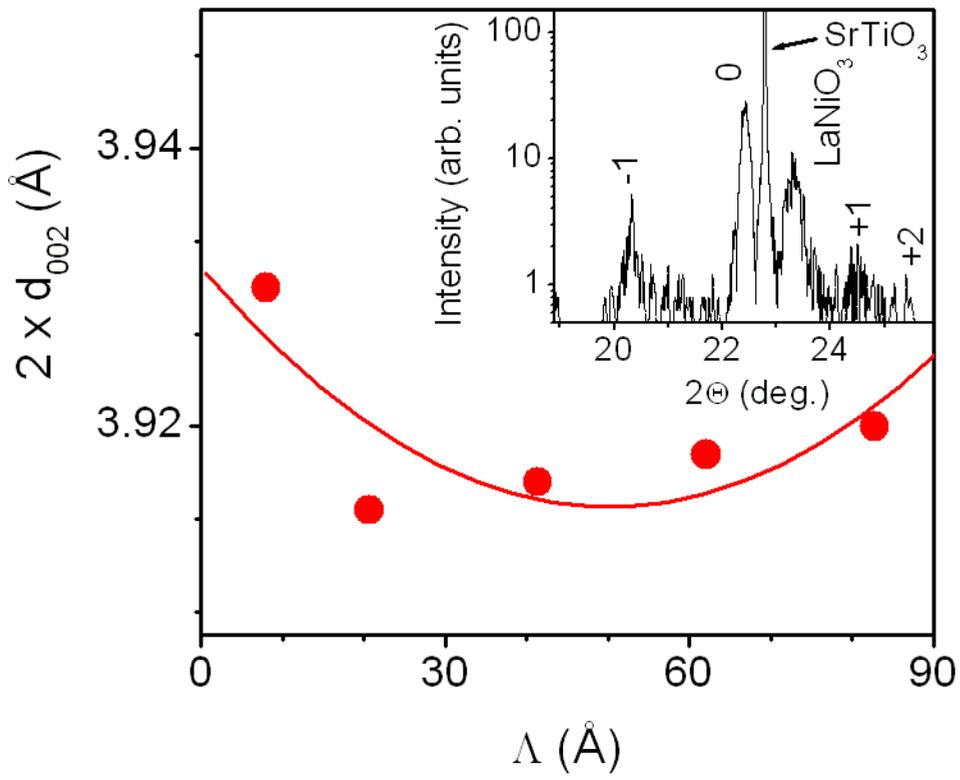

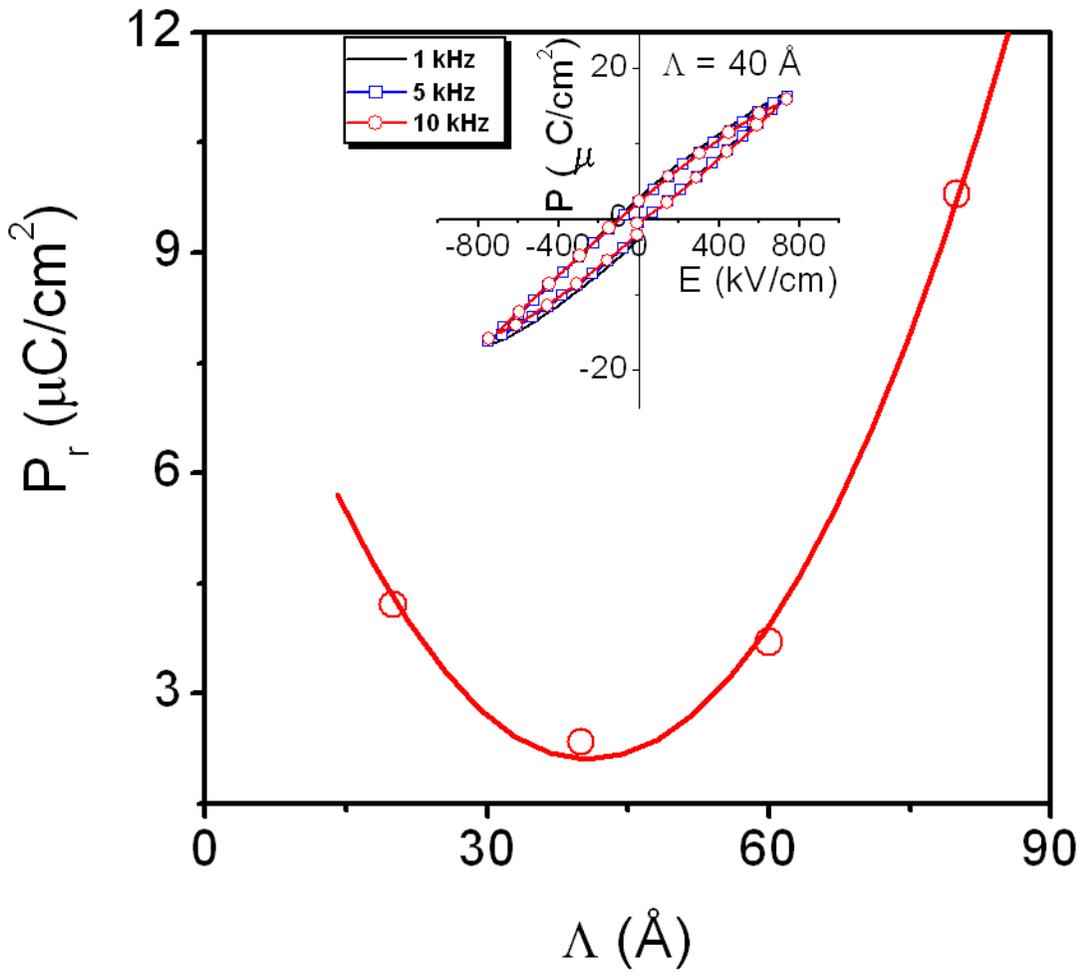

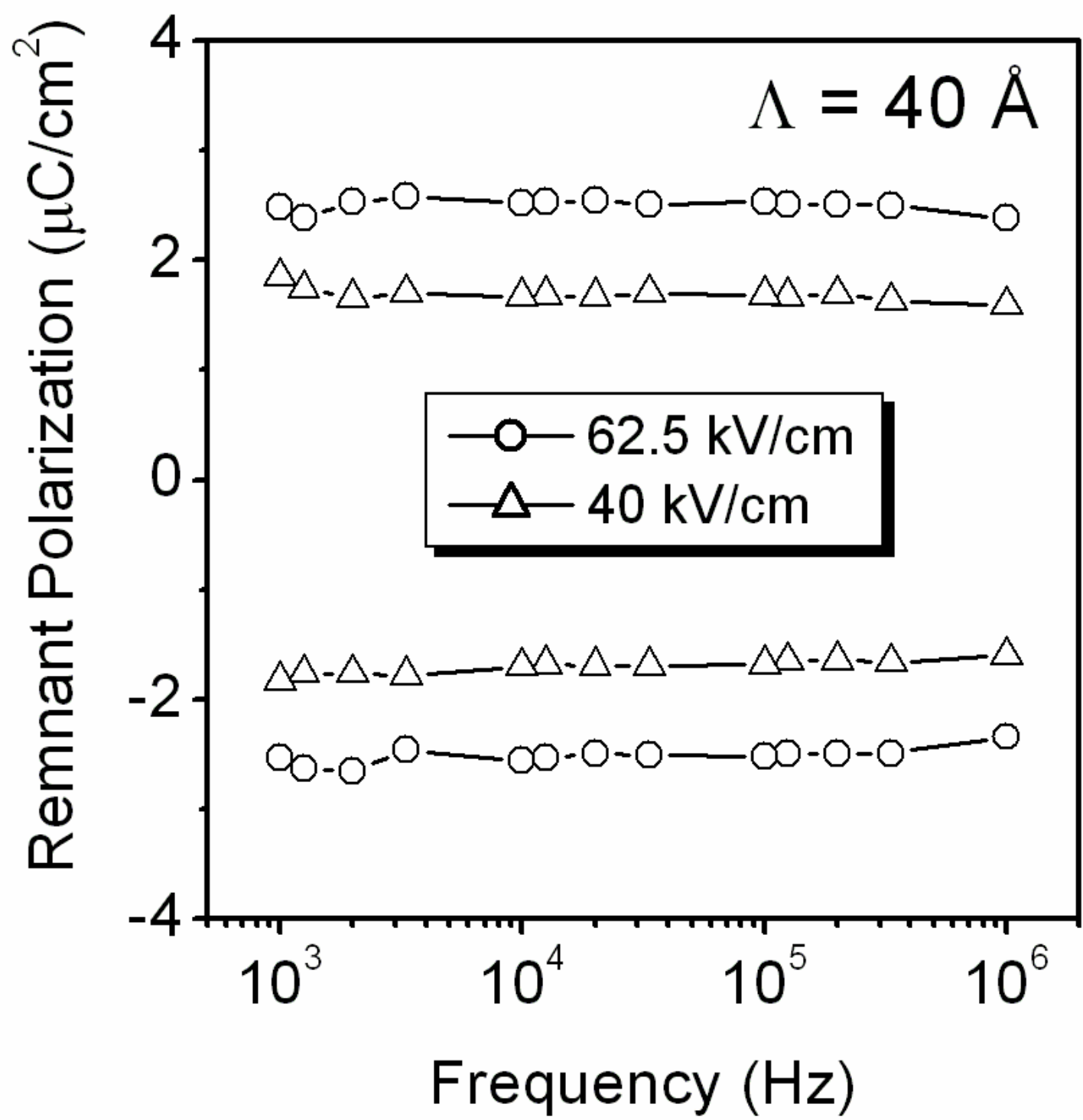

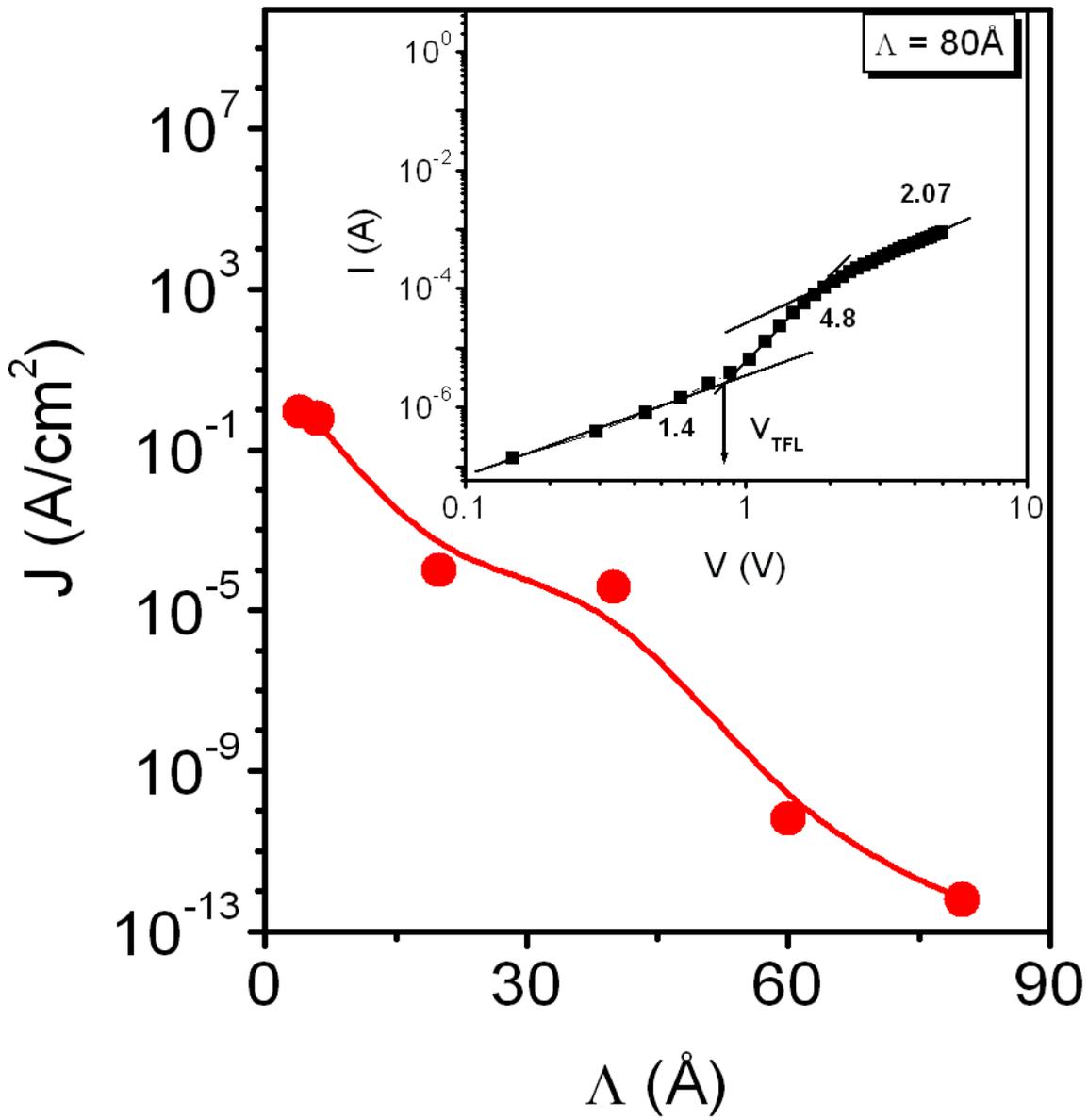